\documentclass[12pt]{article}

\usepackage[table]{xcolor}
\usepackage{amsmath,amssymb,amsthm,algpseudocode,array,booktabs,bm,cancel,cellspace,placeins}
\usepackage{enumerate,graphicx,adjustbox}
\usepackage{pdflscape,mdframed,multirow,natbib,setspace,subcaption,verbatim}
\allowdisplaybreaks

\usepackage[retainorgcmds]{IEEEtrantools}
\usepackage[top=2.54cm, bottom=2.54cm, left=2.54cm, right=2.54cm]{geometry}
\usepackage[flushleft]{threeparttable}
\usepackage{pdflscape}
\usepackage{pgfplots}

\usepackage{xargs}
\usepackage[colorinlistoftodos,prependcaption,textsize=tiny]{todonotes}

\usepackage{hyperref}
\usepackage{chngcntr}

\pgfplotsset{compat=1.15}
\begin{document}

\setlength{\abovedisplayskip}{2pt}
\setlength{\belowdisplayskip}{2pt}

\title{Hot Days, Unsafe Schools? The Impact of Heat on School Shootings} \vspace{0.4in}

\author{
Seunghyun Lee\thanks{Department of Agricultural and Resource Economics, North Carolina State University. Email: slee229@ncsu.edu}
\and
Goeun Lee\thanks{Department of Agricultural and Resource Economics, University of California, Davis. Email: gonlee@ucdavis.edu}
}
\date{}
\maketitle

\vspace{0.5cm}

\pagenumbering{arabic}

\renewcommand*{\theHsection}{chX.\the\value{section}}
\begin{singlespace}
\begin{abstract}
Using data on shootings in U.S.\ K--12 schools from 1981 to 2022, we estimate the effect of temperature on school shootings and assess climate-change impacts. We find that days with maximum temperatures above 90$^{\circ}$F increase school shooting incidence by approximately 90\% relative to days with maximum temperatures below 70$^{\circ}$F. The response is concentrated in interpersonal incidents and in non-class periods, such as before school, dismissal, after school, and lunch: shootings during these periods more than triple on days with maximum temperatures above 90$^{\circ}$F, while shootings during class time show no detectable temperature response. The estimated effects are positive for both indoor and outdoor shootings and are larger for shootings involving fatalities or injuries than for shootings involving only minor or no injuries. Applying the estimated dose-response to future warming, we estimate that interpersonal school shootings increase by 6\% by mid-century (2051--2060) under moderate emissions (SSP2--4.5) and 8\% under high emissions (SSP5--8.5), or about 12 and 16 additional incidents per decade. The present discounted value of mid-century social costs is \$599 million under SSP2--4.5 and \$799 million under SSP5--8.5, driven primarily by lost lifetime earnings among exposed students. The results suggest that climate damages in schools may include rare but high-cost safety events, not only heat stress and learning losses.
\end{abstract}
\end{singlespace}


\newpage
\section{Introduction}

More than 50 million students attend K--12 schools in the United States each day, making schools a central setting for student exposure to extreme heat. Existing work shows that heat reduces learning and cognitive performance among students~\citep{park2020heat, park2022hot, graff2018temperature}, but much less is known about whether heat also affects gun violence in schools. School shootings are rare, but their incidence has risen sharply, roughly doubling over the past two decades~\citep{CHDS}. A large body of evidence shows that high temperatures increase violent crime in the general population~\citep{ranson2014crime, heilmann2021urban, cohen2024understanding, colmer2023access}. Whether and to what extent this relationship extends to schools is an open empirical question, because exposure, adaptation, and supervision differ from those in most general-crime settings. This relationship is relevant for school adaptation decisions because cooling investments, schedule changes, and supervision policies may affect safety as well as learning. In this paper, we ask the following questions: Does extreme heat increase K--12 school shootings, and to what extent? Do the effects differ by incident type, injury severity, incident location, and timing within the school day? Are the timing patterns consistent with adult supervision moderating the heat--violence relationship? And how does future warming affect the incidence and social costs of school shootings?

Schools differ from most settings studied in the heat and crime literature in ways that matter for both exposure and adaptation. Students cannot easily avoid heat by changing when or whether they attend school. Attendance is compulsory and school calendars are fixed. Adaptation through cooling is also limited. During our 1981--2022 sample period, air conditioning was not universal in U.S.\ classrooms, especially outside historically hot regions~\citep{park2020heat}. The daily structure of schools creates within-day variation in supervision: class periods are closely monitored, while lunch, arrival, dismissal, and other transition periods provide less structured oversight. The costs of an incident extend beyond the immediate victims. Students exposed to school shootings suffer long-run losses in mental health, educational attainment, and earnings~\citep{cabral2021trauma}. Many remain enrolled at the affected school after the incident, limiting avoidance even after exposure~\citep{cabral2021trauma}.

We use data from the K--12 School Shooting Database (K--12 SSDB)~\citep{CHDS}, which records both fatal and non-fatal gun incidents in U.S.\ schools. Non-fatal events account for 79\% of school gun incidents in our data. This coverage is important because exposure to non-fatal school shootings also reduces long-term earnings and educational attainment~\citep{cabral2021trauma}, while most existing studies rely on datasets that record only fatal events~\citep{soni2023mass}. We aggregate incidents to daily county-level counts and estimate the effects of temperature using a Poisson quasi-maximum likelihood (PQML) framework. The design uses day-to-day variation in local maximum temperature, controlling for precipitation, state-by-year fixed effects, county-by-day-of-year fixed effects, and day-of-week fixed effects. The estimates therefore compare hotter and cooler days within the same county and time of year, net of state-year shocks and systematic weekday patterns.

We find that days with maximum temperatures above 90$^{\circ}$F increase the incidence of K--12 school shootings by approximately 90\% relative to days with maximum temperatures below 70$^{\circ}$F. The estimates differ sharply by timing. Incidents in non-class periods, such as before school, dismissal, after school, and lunch, more than triple on days with maximum temperatures above 90$^{\circ}$F, while incidents during class time show no detectable temperature response. The estimated high temperature effects are positive for both indoor and outdoor shootings. They are also much larger for interpersonal incidents---those arising from disputes, threats, or targeted violence---than for accidental discharges or suicides, and larger for incidents involving fatalities or injuries than for incidents involving only minor or no injuries. The timing, type, and severity results are consistent with heat increasing the risk of interpersonal gun violence, especially outside structured class periods.

We then apply the estimated dose-response to future climate exposure to assess climate-change impacts on interpersonal school shootings. Under moderate emissions (SSP2--4.5), warming increases interpersonal shootings by 6\% by mid-century (2051--2060), or about 12 additional incidents per decade. The effect rises to 9\% by end-of-century. As a point of comparison, \citet{ranson2014crime} estimates a 3.6\% late-century increase in murder under a comparable emissions pathway, though school shootings are much less frequent than murders. We estimate that the present discounted value of social costs over the mid-century decade is \$599 million (2025 dollars), accounting for lost earnings, mortality, and medical costs. Because this calculation excludes several costs, including psychological costs borne by families, teachers, and surrounding communities, it should be interpreted as a lower-bound estimate.

The results connect three lines of work that are often studied separately: temperature and crime, heat and education, and the measurement of climate damages. First, the paper contributes to the temperature and crime literature by studying a setting where exposure and supervision are institutionally determined. Recent work has moved beyond average heat effects to the channels and moderators underlying the relationship, including shifts in time use and alcohol consumption~\citep{cohen2024understanding}, direct effects on aggression~\citep{mukherjee2021causal, almaas2019destructive}, and institutional factors such as gun laws~\citep{colmer2023access}. The school setting provides within-day variation in supervision, which we use to examine whether the heat--violence relationship is strongest when supervision is limited. The timing pattern and concentration among interpersonal incidents are consistent with adult supervision moderating whether interpersonal disputes result in shootings.

Second, the paper shows that the costs of heat in schools extend beyond learning. Existing studies show that high temperatures reduce cognitive performance and academic achievement~\citep{park2020heat, park2022hot, graff2018temperature}, with similar patterns in developing countries~\citep{garg2020temperature, fishman2019long}. We document a separate safety channel: high temperatures also increase the risk of gun violence in schools. This changes the benefit-cost calculation for school-wide heat adaptation. Investments such as cooling infrastructure or schedule changes should be evaluated not only against avoided learning losses, but also against the potential reduction in rare, high-cost shootings induced by heat.

Third, the paper quantifies a previously omitted component of climate damages: heat-induced school shootings. This channel is costly because school shootings affect not only direct victims but also exposed students. \citet{cabral2021trauma} show that each exposed student loses approximately \$129{,}000 in lifetime earnings (2025 dollars). In our social-cost calculations, costs to exposed students account for over 95\% of the per-incident social cost. This occurs because each shooting exposes many students beyond those who are physically injured, and prior work estimates large long-run earnings losses for exposed students. Existing climate-crime damage estimates generally value direct crime outcomes such as mortality, injury, or victimization, so they do not capture these exposure costs borne by students~\citep{ranson2014crime}. We provide the school-shooting dose-response needed to incorporate these costs into estimates of climate damages.

Section~\ref{sec:background} reviews the background. Sections~\ref{sec:data} and~\ref{sec:research_design} describe the data and empirical strategy. Section~\ref{sec:results} presents results, robustness checks, and the climate-change impact assessment. Section~\ref{sec:conclusion} concludes.

\FloatBarrier
\section{Background}
\label{sec:background}
The literature on heat and crime shows that higher temperatures are associated with more conflict and violence~\citep{burke2015climate}. Prior work points to two mechanisms that are relevant for schools. First, hot weather can change time use and social interaction in public settings, increasing opportunities for conflict. \citet{cohen2024understanding} show that this margin explains part of the heat-crime relationship. Second, heat may affect aggression or self-control among people who are already exposed to one another.

Schools differ from the public settings often studied in the heat-crime literature because exposure is structured. Enrollment, staffing, and daily schedules are largely fixed, and class cancellations due to hot weather are generally uncommon in the U.S.~\citep{alberto2021too}. This does not eliminate all exposure margins, since absences and within-school movement may still vary. It does, however, limit explanations based only on changes in public activity or the volume of spontaneous social interactions. Our empirical goal is not to separately identify each mechanism. Instead, these mechanisms motivate two heterogeneity analyses: whether effects differ by incident type and whether they differ across more and less structured parts of the school day.

Prior evidence on heat and behavior motivates the incident-type comparison. In a laboratory experiment, \citet{almaas2019destructive} find more destructive behavior among participants exposed to thermal stress. In correctional facilities, \citet{mukherjee2021causal} document that days exceeding 80$^{\circ}$F are associated with a 20\% increase in violent interactions and an 18\% increase in violent incidents among incarcerated people. Heat may also affect the behavior of both parties to a confrontation. \citet{annan2023hot}, for example, find suggestive evidence that police shootings rise on days with higher temperatures partly because suspects behave more aggressively. In schools, this evidence suggests that any temperature response should be larger for interpersonal shootings---incidents arising from disputes, threats, or targeted violence---than for accidental discharges or suicides.

The structure of the school day motivates the timing comparison. Adult oversight varies across the day: classrooms are closely monitored, while arrival, dismissal, lunch, and other transition periods are less structured. If supervision reduces the likelihood that conflict results in a shooting, temperature effects should be concentrated outside class periods rather than during class time. We test this prediction by comparing effects across these parts of the school day.

\FloatBarrier
\section{Data}
\label{sec:data}
Our estimation sample is a balanced county-day panel for the contiguous United States from 1981 through 2022. The outcome data come from the K--12 School Shooting Database (K--12 SSDB)~\citep{CHDS}.\footnote{The 2023 version used in this paper was publicly available through the Naval Postgraduate School Center for Homeland Defense and Security at \url{https://www.chds.us/ssdb/} when we downloaded the data. A current version of the database is now hosted at \url{https://k12ssdb.org/}, where users can request access to the data.} We aggregate K--12 SSDB records to daily counts for each county and assign zero shootings to county-days with no recorded incident. The resulting panel contains daily observations for 3,107 counties over 42 years.

The K--12 SSDB defines a school shooting broadly as any incident in which a gun is fired, a gun is brandished or pointed at a person with intent, or a bullet strikes school property. The definition applies regardless of the number of victims, the timing of the incident, or the underlying motive. This broad definition captures a wider range of gun-related incidents in K--12 school settings than datasets limited to fatal or mass shootings. The K--12 SSDB compiles incidents from peer-reviewed studies, government reports, media outlets, and non-profit organizations.

This coverage is important for our setting because most school gun incidents are non-fatal. Fatal shootings account for 21\% of all K--12 SSDB incidents in our sample, meaning that fatal-only datasets omit most events. For example, the Mother Jones database includes only shootings with three or more fatalities~\citep{soni2023mass}, and the Report on School-Associated Violent Deaths from the National School Safety Center also focuses on fatal shootings. Non-fatal incidents are more frequent, and exposure to such incidents can reduce long-term earnings and educational attainment among students~\citep{cabral2021trauma}.

We restrict the analysis to incidents that occur during regular school-day operations. Of the 2,067 geocoded K--12 SSDB incidents in counties in the contiguous United States during 1981--2022, 1,360 occur during regular school-day operations. Specifically, we keep weekday incidents (Monday through Friday) whose K--12 SSDB \texttt{Time\_Period} field is recorded as Morning Classes, Afternoon Classes, Lunch, Dismissal, School Start, Before School, or After School.\footnote{The K--12 SSDB labels are coder-assigned and reflect both clock time and school-activity context. In our data, Dismissal incidents cluster at 2--3~PM (median 3~PM), After School at 3--5~PM (median 4~PM), Evening at 6--9~PM (median 7~PM), and Night at 9~PM--2~AM. The kept buckets correspond to periods when students are likely to be on school grounds. Evening and Night labels apply when the school day has ended. We additionally include weekday incidents whose \texttt{Time\_Period} is missing or marked ``Unknown'' but whose \texttt{During\_Classes} flag indicates the shooting occurred during class time. Narrative review confirms these are daytime weekday events.} We exclude incidents that occurred in the evening or at night, on weekends, or on non-school days such as snow days, holidays, and summer recess, as well as those recorded as occurring during sport events or other school events, which typically fall outside regular instructional time.

\begin{table}[!htbp]
\centering
  \caption{Summary Statistics: Dependent Variables}

\label{tab:sumstat}
\begin{adjustbox}{width=0.7 \textwidth}

\begin{tabular}{@{\extracolsep{5pt}}lcc}
\hline
 & \multicolumn{1}{c}{\shortstack{Mean \\ (per 10,000 county-days)}} & \multicolumn{1}{c}{N} \\
\hline
\multicolumn{3}{l}{\textbf{All Types of Shooting}} \\
All Shooting & 0.285 & 47,676,720 \\
\midrule
\multicolumn{3}{l}{\textbf{Type}} \\
Interpersonal & 0.167 & 47,676,720 \\
Non-Interpersonal & 0.087 & 47,676,720 \\
Missing & 0.031 & 47,676,720 \\
\midrule
\multicolumn{3}{l}{\textbf{Injury Severity}} \\
Fatal or Wounded & 0.165 & 47,676,720 \\
Minor or No Injury & 0.121 & 47,676,720 \\
\midrule
\multicolumn{3}{l}{\textbf{Timing of Incident}} \\
Non-Class Time & 0.142 & 47,676,720 \\
Class Time & 0.143 & 47,676,720 \\
\midrule
\multicolumn{3}{l}{\textbf{Location of Incident}} \\
Indoor & 0.117 & 47,676,720 \\
Outdoor & 0.164 & 47,676,720 \\
Miscellaneous or Missing & 0.005 & 47,676,720 \\
  \bottomrule
\hline
\end{tabular}

\end{adjustbox}

\begin{tablenotes}
    {\footnotesize \item \textit{Notes:} The table presents summary statistics of key dependent variables. All values in the ``Mean'' column are expressed as incidence rates per 10{,}000 county-days. ``All Types of Shooting'' shows the total number of shooting incidents in a given county on a given day. Interpersonal incidents are those classified as ``Anger Over Grade/Suspension/Discipline,'' ``Escalation of Dispute,'' ``Domestic w/ Targeted Victim,'' ``Bullying,'' ``Indiscriminate Shooting,'' ``Drive-by Shooting,'' ``Racial,'' ``Self-defense,'' ``Hostage/Standoff,'' ``Officer-Involved Shooting,'' ``Murder/Suicide,'' and ``Murder/Assassination'' in the K--12 SSDB, whereas non-interpersonal incidents are those that fall into ``Suicide/Attempted,'' ``Accidental,'' ``Illegal Activity,'' ``Intentional Property Damage,'' or ``Psychosis.'' The analysis sample restricts to incidents that occur during normal school operations: weekday daytime events with Time\_Period in a school-day bucket (Morning Classes, Afternoon Classes, Lunch, Dismissal, School Start, Before School, After School). Sport events, evening and night incidents, weekend events, and incidents on non-school days are excluded.}
\end{tablenotes}
\end{table}

Table~\ref{tab:sumstat} presents summary statistics for the outcome variables. School shootings occur at an average rate of 0.285 incidents per 10{,}000 county-days, indicating that the outcome is rare.

The K--12 SSDB classifies each incident by situation, which we group into interpersonal and non-interpersonal categories.\footnote{We define interpersonal incidents as those involving ``Anger Over Grade/Suspension/Discipline,'' ``Escalation of Dispute,'' ``Domestic w/ Targeted Victim,'' ``Bullying,'' ``Indiscriminate Shooting,'' ``Drive-by Shooting,'' ``Racial,'' ``Self-defense,'' ``Hostage/Standoff,'' ``Officer-Involved Shooting,'' ``Murder/Suicide,'' and ``Murder/Assassination'' in the K--12 SSDB.} Interpersonal incidents account for the largest share (0.167 cases per 10{,}000 county-days), followed by non-interpersonal incidents (0.087), which include suicidal, accidental, or property-targeting incidents. Non-interpersonal incidents consist primarily of accidental discharges (42\%), suicides or suicide attempts (31\%), and illegal activity such as drug-related incidents (14\%), with the remainder involving intentional property damage and psychosis-related events. Accidental discharges---the largest subcategory---typically involve a student bringing a firearm to school and the weapon firing unintentionally during handling, often while being shown to peers. Over 90\% involve only a single shot fired.

We classify injury severity based on whether any victim in the K--12 SSDB victim records sustained a fatal or wounded injury. Incidents resulting in fatal or wounded victims occur more frequently than those involving only minor or no injuries (0.165 versus 0.121 cases per 10{,}000 county-days). Incidents classified as minor or no injury are not necessarily benign: 69\% involve shots fired that do not strike anyone, and 10\% involve a gun brandished but not discharged. In roughly 15\% of these cases, narratives describe explicit intervention by teachers, school resource officers, or bystanders---such as tackling the shooter, wrestling a weapon away, or talking an armed individual into surrendering.\footnote{These shares exclude 126 suicide or attempted suicide incidents that fall into the minor-or-no-injury category because the shooter's death is not recorded in the victim injury data.}

By timing, shootings occur during class time and outside class periods (before or after class or during lunch) at similar rates (0.143 versus 0.142 cases per 10{,}000 county-days). By location, more school shootings take place outdoors than indoors (0.164 versus 0.117 cases per 10{,}000 county-days). Among outdoor incidents, parking lots are the most common location (35\%), followed by the front of the school building (21\%) and areas beside the building (16\%). The K--12 SSDB records the location where the shooting occurred, not where the initial conflict or triggering event began. A dispute that begins inside a classroom but results in a shooting in a parking lot is therefore classified as an outdoor incident, and vice versa. This distinction may still inform allocating security resources across indoor and outdoor settings.

We source county-level, population-weighted daily measures of maximum temperature and precipitation from \citet{lee_prism} for the estimation period (1981--2022). These measures are constructed using 4 km--resolution gridded observations produced by the PRISM Climate Group. Table~\ref{tab:sumstat_indep} summarizes the weather variables used in the analysis. County-days with maximum temperatures below 70$^{\circ}$F account for 50.8\% of the sample, while county-days with maximum temperatures above 90$^{\circ}$F account for 10.0\%. Appendix Figure~\ref{fig:temp} maps average daily maximum temperatures and the number of days with maximum temperatures above 90$^{\circ}$F per year across counties. Counties in Texas, Arizona, and parts of Florida have the highest average maximum temperatures and the most frequent exposure to maximum temperatures above 90$^{\circ}$F.

\begin{table}[!htbp]
\centering
  \caption{Summary Statistics: Independent Variables}

\label{tab:sumstat_indep}
\begin{adjustbox}{width=0.5\textwidth}

\begin{tabular}{@{\extracolsep{5pt}}lccc}
\hline
 & \multicolumn{1}{c}{Mean} & \multicolumn{1}{c}{Std.\ Dev.} & \multicolumn{1}{c}{N} \\
\hline
  \multicolumn{4}{l}{\textbf{Temperature ($^{\circ}$F)}} \\
Below 70$^{\circ}$F & 0.508 & 0.500 & 47,676,720 \\
70 - 80$^{\circ}$F & 0.184 & 0.388 & 47,676,720 \\
80 - 90$^{\circ}$F & 0.208 & 0.406 & 47,676,720 \\
Above 90$^{\circ}$F & 0.100 & 0.300 & 47,676,720 \\
  \midrule
  \multicolumn{4}{l}{\textbf{Precipitation (Inch)}} \\
0 inch & 0.525 & 0.499 & 47,676,720 \\
0-0.5 inch & 0.409 & 0.492 & 47,676,720 \\
0.5-1 inch & 0.044 & 0.204 & 47,676,720 \\
Above 1 inch & 0.023 & 0.149 & 47,676,720 \\
\hline \\[-1.8ex]
\end{tabular}

\end{adjustbox}

\begin{tablenotes}
{\footnotesize \item \textit{Notes:} The table presents summary statistics of weather variables used as independent variables. Each variable is a binary indicator for the corresponding category. Temperature refers to the daily maximum, while precipitation denotes total daily rainfall in a given county.}
\end{tablenotes}
\end{table}

\begin{figure}[!htbp]
    \centering
\begin{subfigure}{.49\textwidth}
  \centering
  \caption{School Shootings by Year}
  \includegraphics[width=\linewidth]{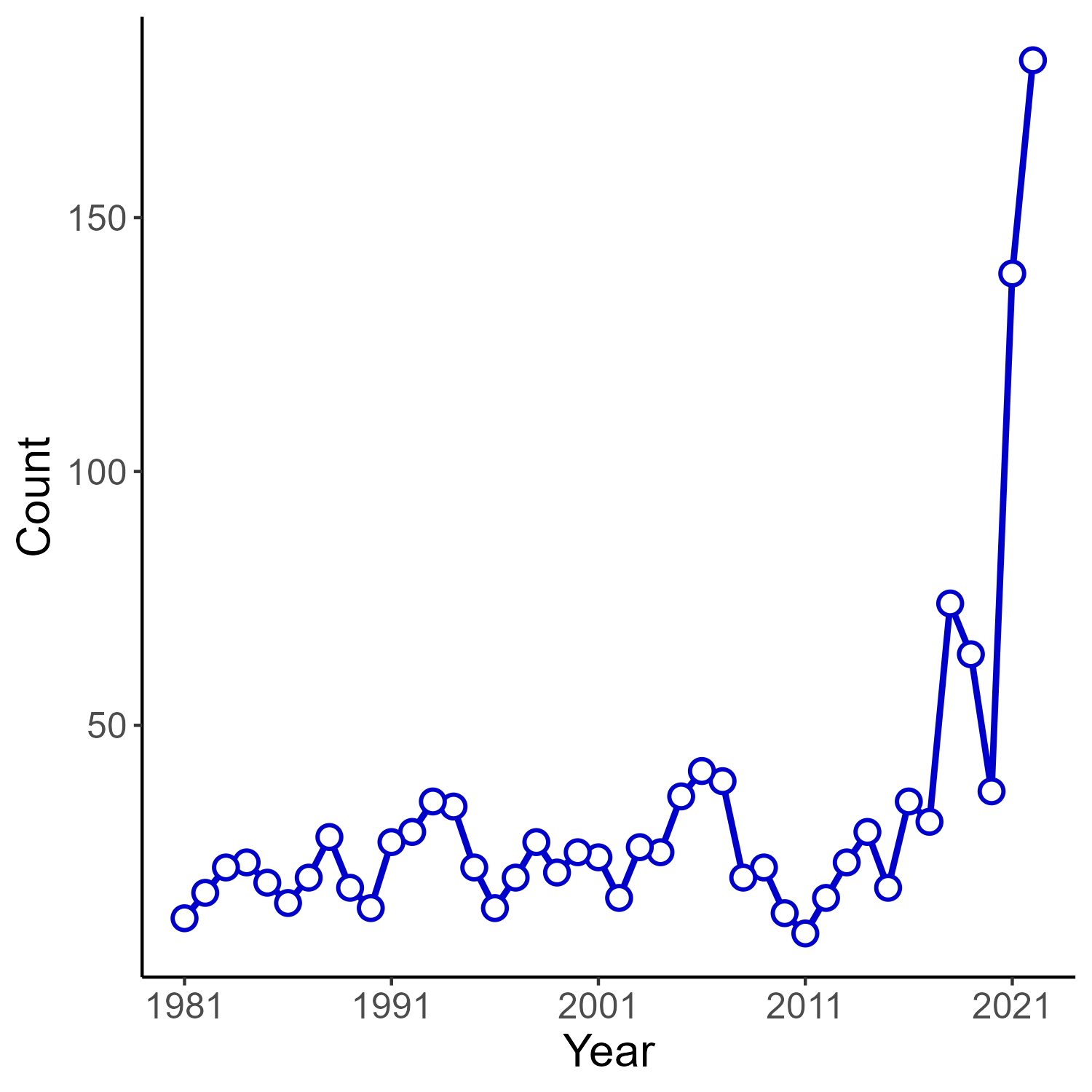} 
\end{subfigure}
\begin{subfigure}{.49\textwidth}
  \centering
    \caption{School Shootings by Month}
  \includegraphics[width=\linewidth]{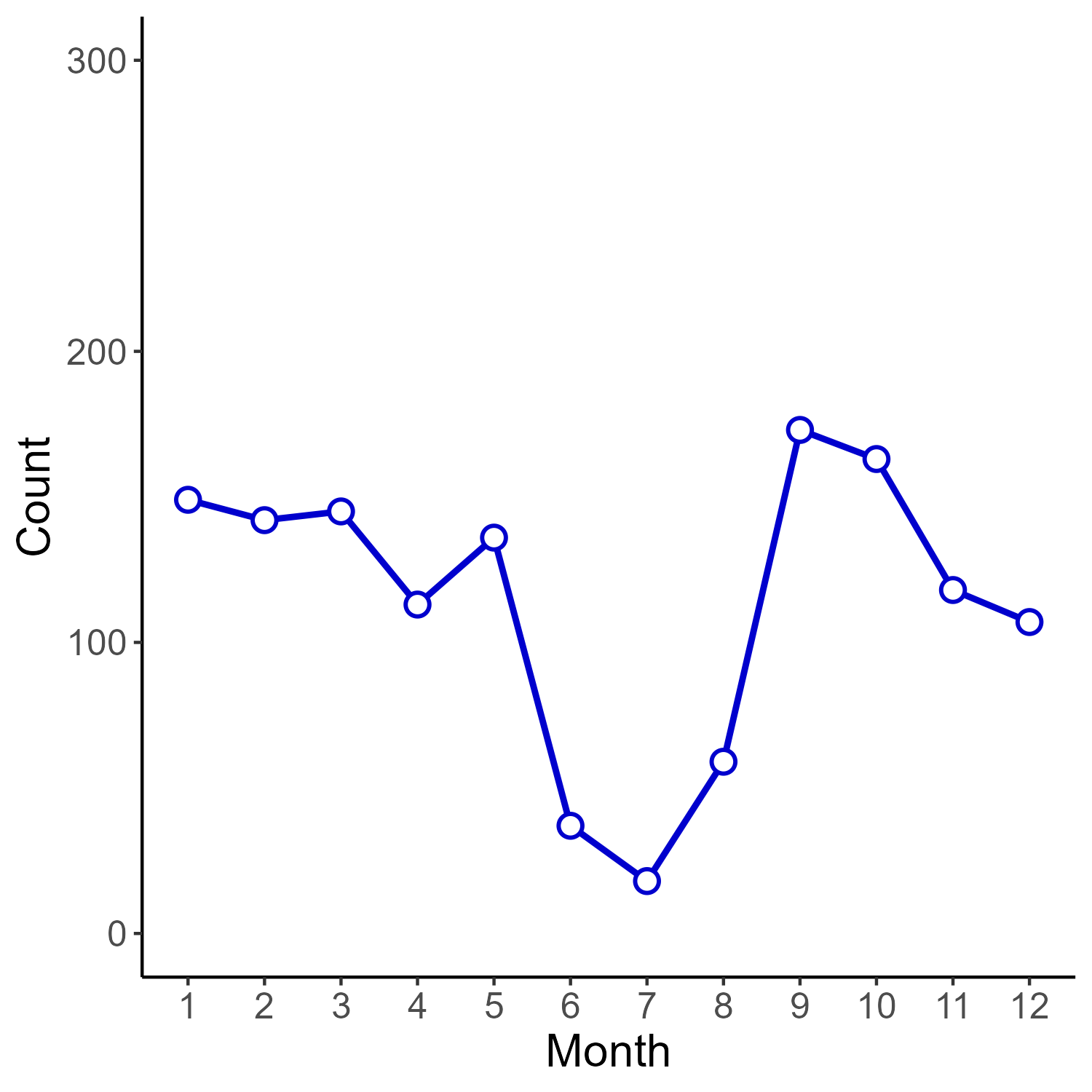}  
  \end{subfigure}
\begin{subfigure}{1\textwidth}
  \centering
     \caption{School Shootings by County}
  \includegraphics[width=\linewidth]{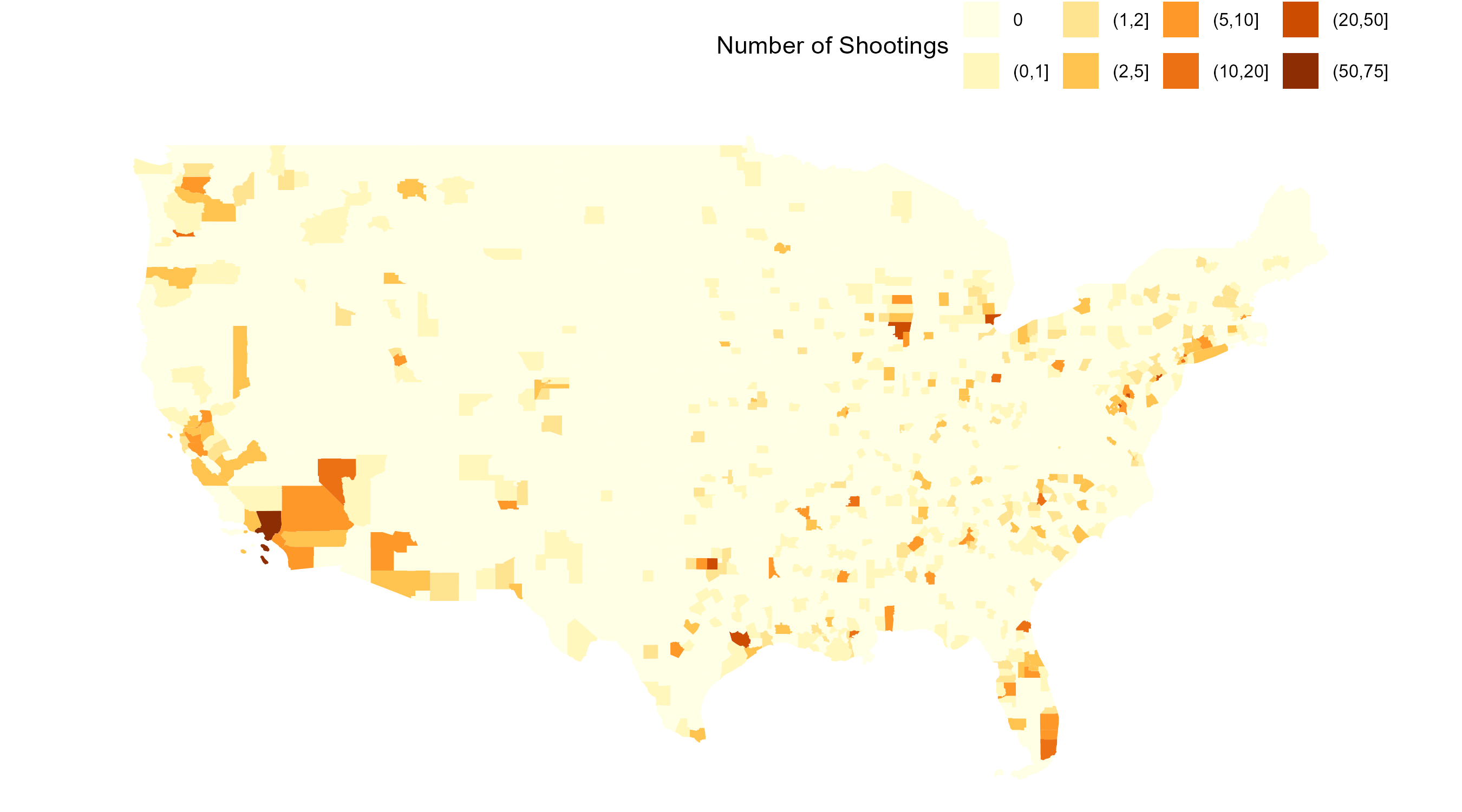}  
\end{subfigure}

      \caption{Temporal and Geographical Variations of School Shootings}
 \begin{minipage}{\textwidth} 
{\footnotesize \textit{Notes}: Panels (a) and (b) in Figure~\ref{fig:shooting} depict the total number of shootings by year and month, respectively. Figure~\ref{fig:shooting} (c) shows the total number of shootings that occurred in each county during the sample period (1981--2022). }
\end{minipage}
\label{fig:shooting}
\end{figure}

Panels (a) and (b) in Figure~\ref{fig:shooting} show the total number of school shootings by year and by month, respectively. The annual number of recorded incidents remained between 10 and 60 until around 2010, after which it increased and reached its sample-period peak in 2021. Panel (c) shows the total number of shootings per county over the sample period. Incidents are less frequent in northern and central counties than in parts of the South, West, and East Coast.

\FloatBarrier
\section{Research Design}
\label{sec:research_design}

We identify the effect of temperature using short-run deviations in daily maximum temperature within counties. The preferred specification compares hotter and cooler days within the same county and calendar day, after absorbing state-by-year shocks and day-of-week patterns. This design follows prior work that uses short-run weather variation at local or individual scales to estimate effects on crime, human capital, and other outcomes~\citep{ranson2014crime, graff2018temperature, heyes2019temperature, heilmann2021urban}. The identifying assumption is that, conditional on precipitation and fixed effects, these short-run temperature deviations are uncorrelated with other determinants of school shootings.

Because school shootings are rare counts with many zeros, we estimate the model using the Poisson quasi-maximum likelihood (PQML) estimator, following \citet{ranson2014crime}, \citet{burkhardt2019effect}, and \citet{bondy2020crime}. PQML models the conditional mean directly, does not require transforming zero outcomes, and remains consistent under correct specification of the conditional mean even if the conditional variance is misspecified~\citep{silva2011further}. Specifically, we parameterize the conditional mean $\mu(X_{cd}) = \mathbb{E}[\text{Shooting}_{cd} \mid X_{cd}]$ using an exponential link:
\begin{equation}
\label{eq:main}
\mu(X_{cd})
= \exp\!\left(
\sum_{k\in K} \beta_k \, \mathbb{I}\!\left(TMAX_{cd} \in k\right)
+ \sum_{l\in L} \delta_l \, \mathbb{I}\!\left(PREC_{cd} \in l\right)
+ \pi_{s\times y}
+ \gamma_{c \times yday}
+ \theta_w
\right).
\end{equation}

We group daily maximum temperature, $TMAX_{cd}$, into 10$^{\circ}$F bins, with
$K \in \{[70^{\circ}\mathrm{F},80^{\circ}\mathrm{F}), [80^{\circ}\mathrm{F},90^{\circ}\mathrm{F}), [90^{\circ}\mathrm{F},\infty)\}$,
and days with maximum temperatures below 70$^{\circ}$F omitted as the reference category.
The coefficients $\beta_k$ are log changes in the expected number of shootings relative to days with maximum temperatures below 70$^{\circ}$F. We report percentage effects as $(\exp(\hat{\beta}_k)-1)\times 100$. To check sensitivity to the reference category, we also disaggregate the below-70$^{\circ}$F bin into 10$^{\circ}$F intervals starting from below 40$^{\circ}$F (Figure~\ref{fig:alternative_bins}). Because school shootings are rare---1{,}360 cases over 42 years---we use broad temperature bins to limit sparsity.
Precipitation, $PREC_{cd}$, is grouped into 0.5-inch intervals up to 1 inch, with
$L \in \{(0,0.5], (0.5,1.0], (1.0,\infty)\}$ (in inches), and zero precipitation serving as the omitted reference category.

The specification includes three sets of fixed effects. State-by-year fixed effects ($\pi_{s\times y}$) absorb annual shocks and policy changes common to counties within the same state. County-by-day-of-year fixed effects ($\gamma_{c\times yday}$) absorb county-specific seasonality, including typical weather patterns and school calendars. Day-of-week fixed effects ($\theta_{w}$) absorb systematic differences across weekdays. Identification therefore comes from deviations in daily temperature from a county's usual temperature on the same calendar day, net of statewide annual shocks, precipitation, and weekday patterns.
We cluster standard errors at the county level to allow arbitrary serial correlation within counties. We report robustness to alternative clustering schemes in Appendix Figure~\ref{fig:cluster}. Regressions are weighted by 2020 county population so that estimates place more weight on counties with larger exposed populations.

\section{Results}
\label{sec:results}
\subsection{Main Results}
\begin{table}[!htbp]
\caption{Temperature Effects on School Shootings (All Types of Shooting)}
\centering
\begin{adjustbox}{width=0.8\textwidth}
\begingroup
\centering
\begin{tabular}{lccc}
   \toprule
                                       & (1)            & (2)           & (3)\\  
   \midrule 
   70-80F                              & 0.0345         & 0.1122        & 0.1210\\   
                                       & (0.0813)       & (0.1113)      & (0.1129)\\   
   80-90F                              & 0.5057$^{***}$ & 0.2690        & 0.4307$^{***}$\\   
                                       & (0.1317)       & (0.1735)      & (0.1591)\\   
   Above 90F                           & 0.4076$^{**}$  & 0.5907$^{**}$ & 0.6557$^{**}$\\   
                                       & (0.1778)       & (0.2750)      & (0.2695)\\   
    \\
   Observations                        & 47,676,720     & 47,676,720    & 47,676,720\\  
    \\
   Precipitation Controls              & $\checkmark$   & $\checkmark$  & $\checkmark$\\   
   County fixed effects                & $\checkmark$   &               & \\  
   Date fixed effects                  & $\checkmark$   &               & \\  
   County x Month x Year fixed effects &                & $\checkmark$  & \\  
   Day of Week fixed effects           &                & $\checkmark$  & $\checkmark$\\   
   County x DOY fixed effects          &                &               & $\checkmark$\\   
   State x Year fixed effects          &                &               & $\checkmark$\\   
   \bottomrule
\end{tabular}
\par\endgroup
\end{adjustbox}

\begin{tablenotes}
{\footnotesize \item \textit{Notes:} Column (3) in Table~\ref{tab:main} presents the regression results from our preferred specification (Equation~\ref{eq:main}), while columns (1)--(2) report results from alternative fixed effects specifications. All models include precipitation controls and are estimated using the Poisson quasi-maximum likelihood estimator. Each coefficient represents the semi-elasticity of school shootings with respect to the indicated bin of daily maximum temperature, relative to the omitted category: days with maximum temperatures below 70$^{\circ}$F. DOY fixed effects refer to day-of-year fixed effects. The regression is weighted using population counts from the 2020 U.S. Census Bureau. Standard errors are clustered at the county level. * $p <$ 0.1. ** $p <$ 0.05. *** $p <$ 0.01.  }
\end{tablenotes}
\label{tab:main}
\end{table}
\FloatBarrier

Table~\ref{tab:main} presents the impact of daily maximum temperatures on school shooting incidence, with different sets of fixed effects used in each column. Across specifications, the two highest temperature bins have positive coefficients. Column (3) reports results from our preferred specification, which includes state-by-year, county-by-day-of-year, and day-of-week fixed effects. We find that on days with maximum temperatures between 80--90$^{\circ}$F and on days with maximum temperatures above 90$^{\circ}$F, the incidence of shootings increases by 53.9\% and 92.7\%, respectively, compared to days with maximum temperatures below 70$^{\circ}$F. Percentage effects are computed as $(\exp(\hat{\beta}) - 1) \times 100$.

Column (1) reports a two-way fixed effects specification with county and date fixed effects. This specification compares counties experiencing different temperatures on the same date, after removing time-invariant county differences and shocks common to all counties on a given day. It is useful as a benchmark, but it does not account for state-specific time-varying shocks, including changes in gun policy, school safety policy, law enforcement, or other state-level conditions, nor does it absorb county-specific seasonality in weather, school calendars, or shooting risk. The estimates for the higher temperature bins are positive. Compared with our preferred specification, the 80--90$^{\circ}$F coefficient is larger and the above-90$^{\circ}$F coefficient is smaller. Column (2) instead includes county-by-month-by-year fixed effects and day-of-week fixed effects. This specification absorbs shocks common to a county in a given month and year, but it may also remove useful temperature variation when hot conditions persist over multiple days or weeks. The above-90$^{\circ}$F coefficient is slightly smaller than the preferred estimate, while the 80--90$^{\circ}$F coefficient is substantially smaller and not statistically significant. Although magnitudes vary across fixed-effect specifications, the estimates provide consistent evidence that higher temperatures increase school shootings.

\subsubsection{Robustness Checks}
\begin{figure}[!htbp]
    \centering
    \includegraphics[width=0.8\linewidth]{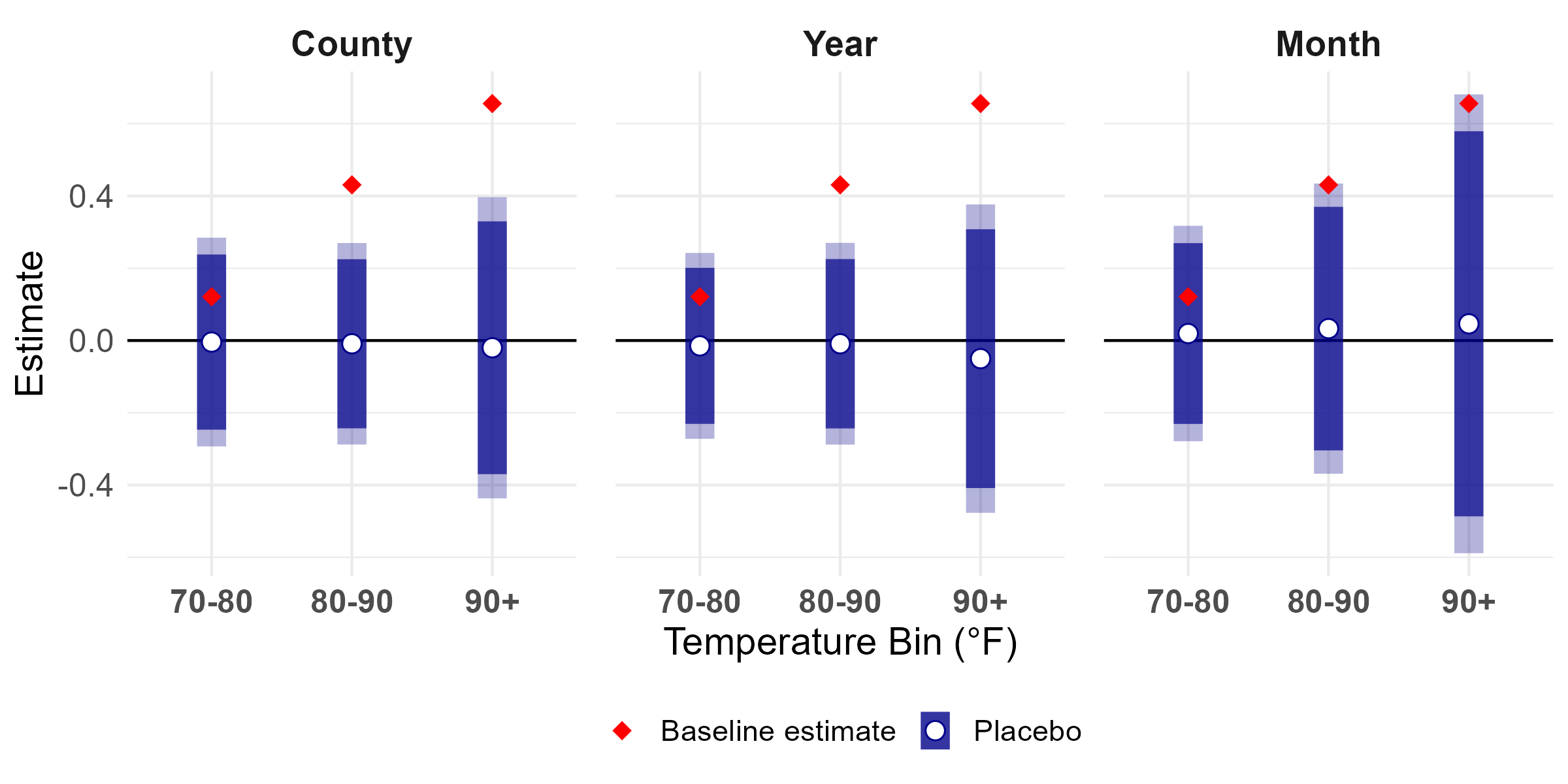}
    \caption{Falsification Tests}
\begin{minipage}{\textwidth}
{\footnotesize \textit{Notes:} Figure~\ref{fig:falsification} presents results from a placebo test. In each iteration, temperature and precipitation bin indicators are jointly reshuffled across observations within the same county (left), year (middle), or month (right), preserving within-day correlations across weather variables. The regression from Equation~\ref{eq:main} is then re-estimated on the reshuffled data. This procedure is repeated 100 times per test. Open circles show the mean placebo coefficient. Thick bars show 90\% and 95\% intervals constructed from the standard deviation of the 100 placebo estimates. Red diamonds show the baseline estimates from the preferred specification (column~3 of Table~\ref{tab:main}).}
\end{minipage}
\label{fig:falsification}
\end{figure}

To assess whether our estimates could arise from chance variation rather than a genuine temperature effect, we conduct placebo tests. We randomly reshuffle temperature and precipitation bin indicators as a block across observations within the same county, year, or month, preserving within-day correlations across weather variables. For each test, we repeat this procedure 100 times and re-estimate Equation~\ref{eq:main} on each reshuffled sample. Figure~\ref{fig:falsification} plots the mean placebo coefficient (open circles) with 90\% and 95\% intervals constructed from the standard deviation of the 100 placebo estimates, alongside the baseline estimates from our preferred specification (red diamonds). Across all three reshuffling dimensions, the placebo coefficients cluster around zero, and the baseline estimates for the 80--90$^{\circ}$F and above-90$^{\circ}$F bins fall well outside the placebo distributions. These exercises suggest that the results are not driven by the placebo variation captured by the reshuffling procedure.

\begin{figure}[!htbp]
    \centering
    \includegraphics[width=0.85\linewidth]{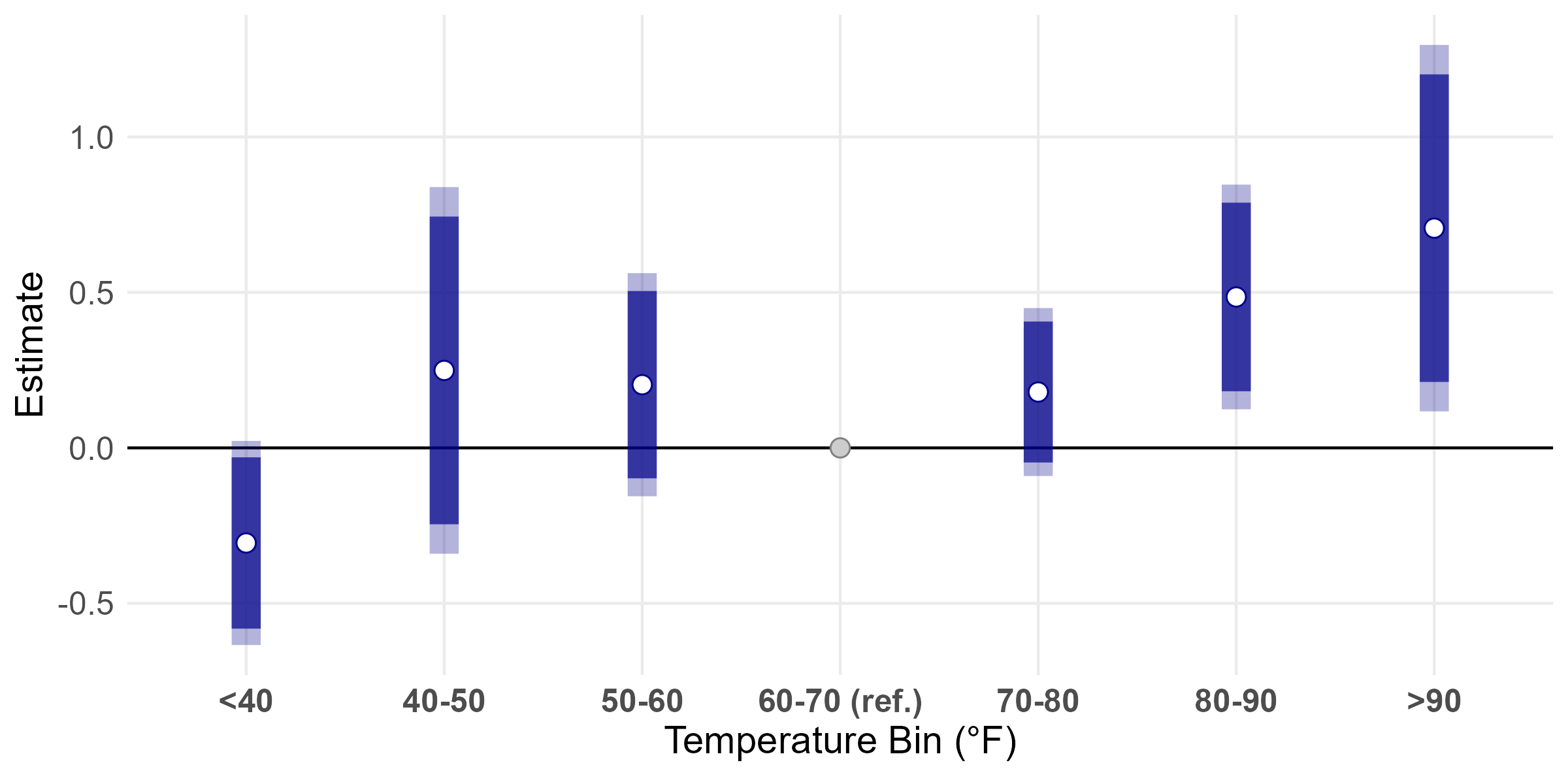}
    \caption{Main Results: Alternative Temperature Bins}
\begin{minipage}{\textwidth}
{\footnotesize \textit{Notes:} Figure~\ref{fig:alternative_bins} plots coefficients from estimating Equation~\ref{eq:main} with finer temperature bins. The omitted category is 60--70$^{\circ}$F (shown in grey at zero). Bars show 95\% confidence intervals constructed from county-clustered standard errors. The regression includes county-by-day-of-year, state-by-year, and day-of-week fixed effects, controls for precipitation, and is weighted using 2020 county population counts from the U.S.\ Census Bureau.}
\end{minipage}
\label{fig:alternative_bins}
\end{figure}

Next, we check whether our results are robust to alternative temperature bins that disaggregate the bin for maximum temperatures below 70$^{\circ}$F. Specifically, we divide the below-70$^{\circ}$F bin into below 40$^{\circ}$F, 40--50$^{\circ}$F, 50--60$^{\circ}$F, and 60--70$^{\circ}$F. Figure~\ref{fig:alternative_bins} plots the coefficients from estimating Equation~\ref{eq:main} with this finer breakdown, using 60--70$^{\circ}$F as the reference category. The estimates for the 80--90$^{\circ}$F and above-90$^{\circ}$F bins remain positive, statistically significant, and similar in magnitude to the baseline estimates in Table~\ref{tab:main}. Coefficients on the colder bins are generally close to zero, except for the below-40$^{\circ}$F coefficient, which is negative and statistically significant at the 10\% level.

We also examine alternative clustering specifications. Appendix Figure~\ref{fig:cluster} presents estimation results when standard errors are clustered by county-year, state, and state-year, in addition to the baseline county-level clustering. The results are similar across these clustering approaches.

\subsection{Heterogeneous Effects of Temperature on School Shootings}
We next assess how the temperature--shooting relationship differs by the timing and location of the event, incident type, and injury severity.

\subsubsection{Heterogeneity by Incident Timing and Location}

\begin{table}[!htbp]
\centering
\caption{Temperature Effects by Incident Timing and Location}
\begin{adjustbox}{width=0.8\textwidth}
\begingroup
\centering
\begin{tabular}{lcccc}
   \toprule
                              & \multicolumn{2}{c}{Timing} & \multicolumn{2}{c}{Location} \\
                              \cmidrule(lr){2-3} \cmidrule(lr){4-5}
                              & Non-Class Time & Class Time    & Indoor        & Outdoor \\   
                              & (1)            & (2)           & (3)           & (4)\\  
   \midrule 
   70-80F                     & 0.1813         & 0.0683        & 0.2983        & 0.0020\\   
                              & (0.1417)       & (0.1442)      & (0.2244)      & (0.1124)\\   
   80-90F                     & 0.8868$^{**}$  & -0.1666       & 0.3587        & 0.4613$^{***}$\\   
                              & (0.3737)       & (0.1424)      & (0.2303)      & (0.1740)\\   
   Above 90F                  & 1.192$^{***}$  & -0.1639       & 0.8448$^{**}$ & 0.6184$^{*}$\\   
                              & (0.3795)       & (0.3102)      & (0.3617)      & (0.3307)\\   
    \\
   Observations               & 47,676,720     & 47,676,720    & 47,676,720    & 47,676,720\\  
    \\
   Precipitation Controls     & $\checkmark$   & $\checkmark$  & $\checkmark$  & $\checkmark$\\   
   State x Year fixed effects & $\checkmark$   & $\checkmark$  & $\checkmark$  & $\checkmark$\\   
   County x DOY fixed effects & $\checkmark$   & $\checkmark$  & $\checkmark$  & $\checkmark$\\   
   Day of week fixed effects  & $\checkmark$   & $\checkmark$  & $\checkmark$  & $\checkmark$\\   
   \bottomrule
\end{tabular}
\par\endgroup
\end{adjustbox}

\begin{tablenotes}
{\footnotesize \item \textit{Notes:} The table presents regression results from our preferred specification (Equation~\ref{eq:main}), estimated separately by incident timing and location. All models include precipitation controls and are estimated using the Poisson quasi-maximum likelihood estimator. Each coefficient represents the semi-elasticity of school shootings with respect to the indicated bin of daily maximum temperature, relative to the omitted category: days with maximum temperatures below 70$^{\circ}$F. DOY fixed effects refer to day-of-year fixed effects. The regression is weighted using population counts from the 2020 U.S. Census Bureau. Standard errors are clustered at the county level. * $p <$ 0.1. ** $p <$ 0.05. *** $p <$ 0.01.}
\end{tablenotes}

\label{tab:timing_location}
\end{table}
\FloatBarrier

We first examine heterogeneity by event timing, distinguishing incidents that occur before or after class or during lunch from those that occur during class time. Columns (1)--(2) of Table~\ref{tab:timing_location} report the results. The temperature effects are concentrated outside class periods. On days with maximum temperatures between 80--90$^{\circ}$F and on days with maximum temperatures above 90$^{\circ}$F, shootings before or after class or during lunch increase by 142.7\% and 229.4\%, respectively. In contrast, class-time point estimates are slightly negative (about $-15\%$ in both temperature bins) and statistically indistinguishable from zero.\footnote{For the 80--90$^{\circ}$F bin, the class-time coefficient is $-0.167$ (SE~=~0.142), and the minimum detectable effect (MDE) at 80\% power is 0.398---well below the non-class estimate of 0.887. For the above-90$^{\circ}$F bin, the class-time coefficient is similarly small ($-0.164$, SE~=~0.310), and the MDE of 0.868 is also well below the non-class estimate of 1.192. The class-time regression therefore has sufficient power to detect an effect of the non-class magnitude in both bins.} This timing pattern is consistent with heat-related shootings being more likely to occur during less structured periods of the school day, when adult oversight is weaker. During class time, structured activities and adult supervision may limit escalation into a shooting.

We also examine whether the effects differ by incident location, distinguishing between indoor and outdoor settings. Incident location does not necessarily correspond to where shooters or victims were exposed to heat stress. Situations may develop in one setting and culminate in another. For example, a conflict that begins indoors may result in a shooting in a parking lot, the most common outdoor incident location in our data. Nonetheless, the location of the incident is relevant for targeting prevention efforts, including the allocation of security resources and supervision. Columns~(3)--(4) of Table~\ref{tab:timing_location} present the results. The estimated high temperature effects are positive for both indoor and outdoor shootings, although statistical significance varies across bins and incident location types. Outdoor shootings increase by approximately 59\% on days with maximum temperatures between 80--90$^{\circ}$F and by about 86\% on days with maximum temperatures above 90$^{\circ}$F, relative to days with maximum temperatures below 70$^{\circ}$F. Indoor shootings increase by roughly 43\% and approximately 133\% in the same two temperature bins. The estimates imply that the temperature response is not limited to outdoor incidents. For prevention, the results do not support focusing heat-related safety planning only on outdoor areas.

\subsubsection{Heterogeneity by Incident Type and Injury Severity}
\begin{table}[!htbp]
\caption{Temperature Effects by Shooting Type and Injury Severity}
\centering
\begin{adjustbox}{width=0.8\textwidth}
\begingroup
\centering
\begin{tabular}{lcccc}
   \toprule
                              & \multicolumn{2}{c}{Type} & \multicolumn{2}{c}{Injury Severity} \\
                              \cmidrule(lr){2-3} \cmidrule(lr){4-5}
                              & Interpersonal  & Non-Interpersonal & Fatal or Wounded & Minor or No Injury \\   
                              & (1)            & (2)               & (3)              & (4)\\  
   \midrule 
   70-80F                     & 0.1541         & 0.1651            & 0.2212           & 0.1007\\   
                              & (0.2180)       & (0.1871)          & (0.2388)         & (0.2932)\\   
   80-90F                     & 0.4444$^{***}$ & 0.5182            & 0.5606$^{***}$   & 0.2886\\   
                              & (0.1705)       & (0.3356)          & (0.2067)         & (0.2128)\\   
   Above 90F                  & 0.7208$^{*}$   & 0.0073            & 0.7696$^{*}$     & 0.4179\\   
                              & (0.3692)       & (0.4548)          & (0.4313)         & (0.4761)\\   
    \\
   Observations               & 47,676,720     & 47,676,720        & 47,676,720       & 47,676,720\\  
    \\
   Precipitation Controls     & $\checkmark$   & $\checkmark$      & $\checkmark$     & $\checkmark$\\   
   State x Year fixed effects & $\checkmark$   & $\checkmark$      & $\checkmark$     & $\checkmark$\\   
   County x DOY fixed effects & $\checkmark$   & $\checkmark$      & $\checkmark$     & $\checkmark$\\   
   Day of week fixed effects  & $\checkmark$   & $\checkmark$      & $\checkmark$     & $\checkmark$\\   
   \bottomrule
\end{tabular}
\par\endgroup
\end{adjustbox}

\begin{tablenotes}
{\footnotesize \item \textit{Notes:} The table presents regression results from our preferred specification (Equation~\ref{eq:main}), estimated separately by shooting type and injury severity. All models include precipitation controls and are estimated using the Poisson quasi-maximum likelihood estimator. Each coefficient represents the semi-elasticity of school shootings with respect to the indicated bin of daily maximum temperature, relative to the omitted category: days with maximum temperatures below 70$^{\circ}$F. DOY fixed effects refer to day-of-year fixed effects. The regression is weighted using population counts from the 2020 U.S. Census Bureau. Standard errors are clustered at the county level. * $p <$ 0.1. ** $p <$ 0.05. *** $p <$ 0.01.}
\end{tablenotes}

\label{tab:type_severity}
\end{table}
\FloatBarrier

We next examine whether temperature effects differ by incident type, classifying shootings as interpersonal or non-interpersonal. Table~\ref{tab:type_severity} presents the results. Interpersonal shootings increase by 55.9\% on days with maximum temperatures between 80--90$^{\circ}$F and by 105.6\% on days with maximum temperatures above 90$^{\circ}$F, both larger than the baseline estimates in Column~(3) of Table~\ref{tab:main}. Column (2) shows no statistically significant temperature effects on non-interpersonal shootings, a category dominated by accidental discharges, illegal activity, and suicides, as described in Section~\ref{sec:data}. The 80--90$^{\circ}$F estimate is positive but imprecise, while the estimate for maximum temperatures above 90$^{\circ}$F is close to zero. We therefore interpret the type results as evidence that the temperature response is concentrated in interpersonal incidents, with the non-interpersonal estimates not statistically distinguishable from zero. The concentration among interpersonal incidents is consistent with evidence from the broader heat-crime literature that higher temperatures increase violent crime~\citep{ranson2014crime, cohen2024understanding, colmer2023access}.

We next examine whether heat affects the severity of shooting outcomes. While the type decomposition classifies incidents by motive, the severity decomposition classifies them by outcome. The two do not map cleanly onto each other: 29\% of interpersonal incidents result in minor or no injury---disputes where shots miss or are fired into the air---and 34\% of non-interpersonal incidents result in fatal or wounded outcomes, such as accidental discharges that wound bystanders. Columns~(3)--(4) of Table~\ref{tab:type_severity} present results by injury severity. Fatal-or-wounded shootings increase by 75\% on days with maximum temperatures between 80--90$^{\circ}$F and by 116\% on days with maximum temperatures above 90$^{\circ}$F, both exceeding the baseline estimates of 54\% and 93\%. Minor-or-no-injury incidents also have positive point estimates, but the estimates are smaller and less precise, especially for days with maximum temperatures above 90$^{\circ}$F. These results suggest that the temperature response is larger for incidents involving fatalities or injuries, though the severity decomposition should be interpreted cautiously. An open question is whether heat generates new severe incidents or escalates incidents that would otherwise have been minor. We cannot statistically separate these margins, but experimental evidence that heat increases aggression~\citep{mukherjee2021causal, almaas2019destructive} is consistent with both.

\subsection{Climate Change Impact Assessments}

This section uses the estimated temperature response to assess how future warming may affect interpersonal school shootings. We focus on interpersonal shootings because the historical estimates show the clearest temperature response for this category and because these incidents are most closely tied to the conflict mechanism discussed above. We apply the interpersonal-shooting estimates from column~(1) of Table~\ref{tab:type_severity}, estimated using the preferred specification, to downscaled CMIP6 climate projections. The exercise holds non-climate determinants of school shootings fixed and then converts the estimated changes in incidents into social cost estimates.

\subsubsection{Climate Change Data}
We use daily climate projections from seven CMIP6 global climate models (GCMs) in the NASA NEX Global Daily Downscaled Projections (NEX-GDDP-CMIP6) archive.\footnote{The CMIP6 models used in this study are GFDL-CM4, GFDL-ESM4, IPSL-CM6A-LR, MIROC6, MPI-ESM1-2-HR, NorESM2-MM, and TaiESM1.} The NEX-GDDP-CMIP6 product provides bias-corrected, spatially downscaled daily climate variables on a 0.25$^{\circ}$ (approximately 25 km) grid for the contiguous United States~\citep{ThrasherEtAl2022}. We average climate variables across the seven models so that the results are not driven by any single climate model.

We use daily maximum temperature and precipitation for three periods: a historical period (2005--2014), a mid-century period (2051--2060), and a late-century period (2091--2100). We consider a moderate-emissions pathway, SSP2--4.5, and a high-emissions pathway, SSP5--8.5. To match the historical weather data used in estimation, we aggregate gridded climate variables to county-level population-weighted means and assign each county-day to the same temperature and precipitation bins used in the main analysis. Appendix Figure~\ref{fig:CMIP6-weight-grid} illustrates the underlying population weights at the native 25~km NEX-GDDP-CMIP6 grid resolution.

\begin{figure}
    \centering
    \includegraphics[width=\linewidth]{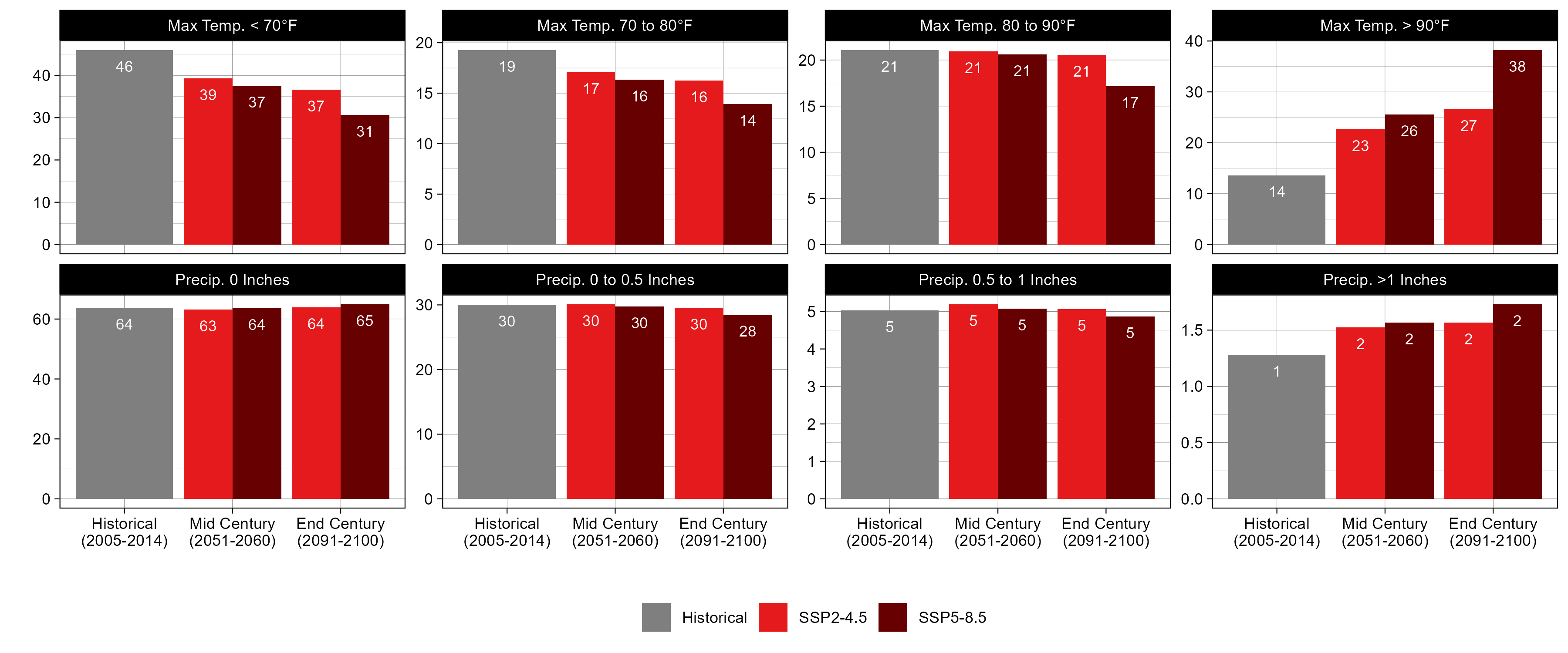}
    \caption{Shifts in Daily Temperature and Precipitation Distributions Under Historical and Future Climate Scenarios}
    \label{fig:bins_by_scenario}
 \caption*{\footnotesize\textit{Notes:} This figure displays the average proportion of days falling into specified temperature and precipitation bins across U.S.\ counties, based on averages across seven downscaled CMIP6 climate simulations. SSP2--4.5 represents a moderate-emissions pathway with emissions stabilizing around mid-century, while SSP5--8.5 represents a high-emissions pathway. Values reflect the mean percentage of days for each scenario and time period.}
\end{figure}

Figure~\ref{fig:bins_by_scenario} summarizes shifts in the distributions of daily maximum temperature and precipitation, averaged across U.S.\ counties and the seven-model ensemble. The share of days with maximum temperatures above 90$^{\circ}$F rises from roughly 14\% in the historical period to 23--26\% by mid-century and 27--38\% by late century, with corresponding declines in milder temperature bins. Changes in precipitation bins are comparatively small.

\begin{figure}
    \centering
    \includegraphics[width=\linewidth]{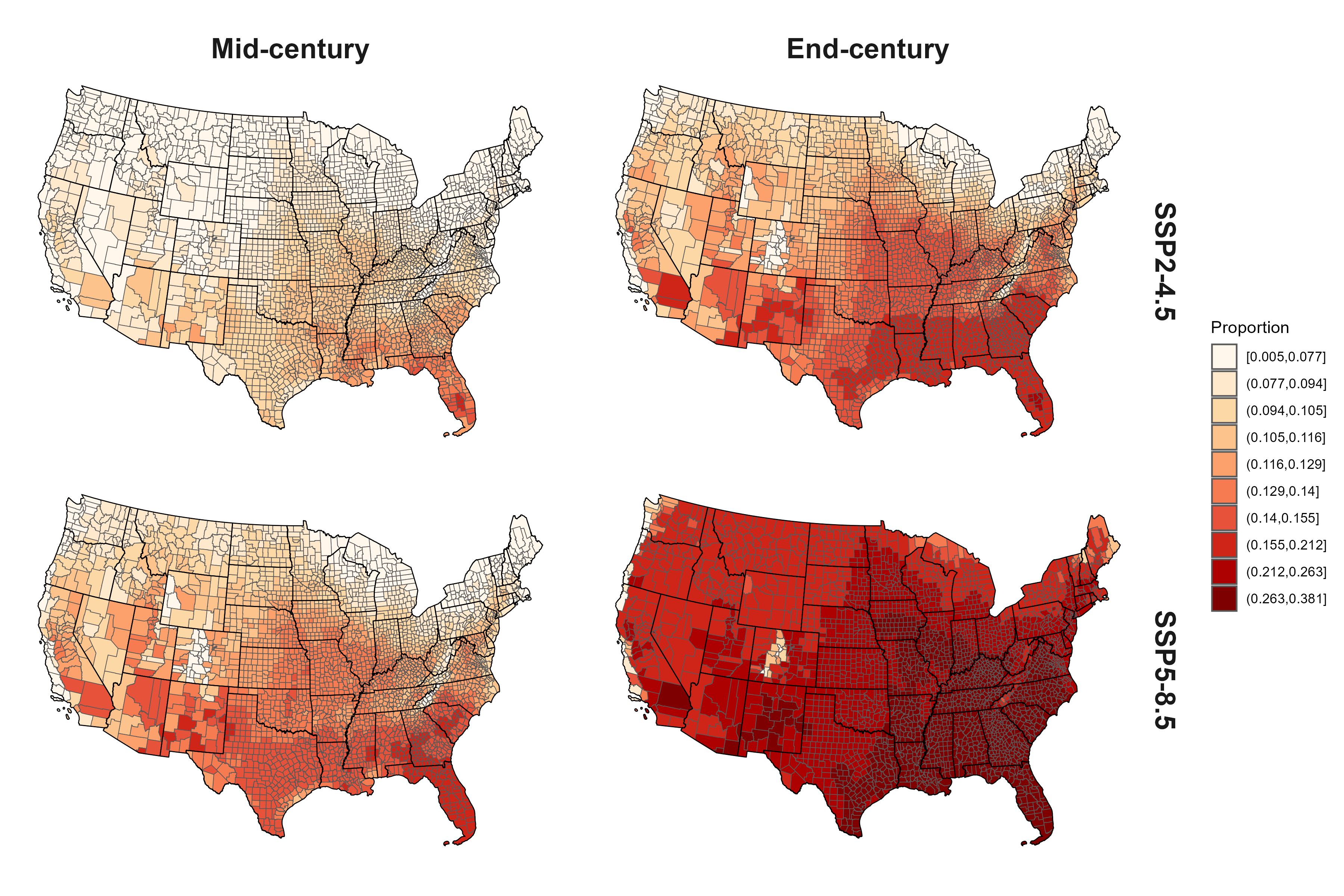}
    \caption{Change in the Frequency of Days with Maximum Temperatures above 90$^{\circ}$F}
   \caption*{\footnotesize\textit{Notes:} Each county-level value represents the ensemble mean across seven downscaled CMIP6 climate models. Changes are computed relative to the baseline period 2005--2014. The mid-century period corresponds to 2051--2060, and the late-century period corresponds to 2091--2100. Results are shown separately for the SSP2--4.5 and SSP5--8.5 scenarios.}

    \label{fig:t90_change}

\end{figure}

Figure~\ref{fig:t90_change} maps changes in the frequency of days with maximum temperatures above 90$^{\circ}$F across counties. By mid-century, increases are concentrated in the Southwest, Gulf Coast, and Southeast, with smaller changes in northern regions. By late century, the increase extends across much of the country, especially under SSP5--8.5.

\subsubsection{Climate Change Impacts on Interpersonal School Shootings}
For each county and scenario, we calculate the change in the distribution of temperature and precipitation bins between the future and historical periods. We then multiply these changes by the estimated coefficients from the interpersonal-shooting specification. The resulting quantity gives the change in log expected shootings attributable to future climate exposure. Exponentiating converts it to a percentage change:

\begin{equation}
\%\Delta \hat{Y}_{c,s} = \left( \exp\left(\Delta W^\prime_{c,s} \hat{\beta} \right) - 1 \right) \times 100\%
\end{equation}

\noindent where $\Delta W_{c,s} = W_{c,s}^{future} - W_{c,s}^{historical}$. The vectors $W_{c,s}^{historical}$ and $W_{c,s}^{future}$ represent the ensemble-average frequencies of temperature and precipitation bins for the historical and future periods, under emissions scenario $s$ in county $c$. We aggregate county-level percentage impacts using 2024 county population weights, although unweighted averages are similar.

Table~\ref{tab:projected_impacts} reports the estimated impacts on interpersonal school shootings. The estimates imply a 6--17\% increase in interpersonal school shootings relative to the 2005--2014 historical climate baseline, with the magnitude varying by emissions pathway and time horizon. By mid-century, interpersonal school shootings increase by 6.37\% under SSP2--4.5 and 8.49\% under SSP5--8.5, corresponding to 12.06 and 16.07 additional incidents per decade. The difference between emissions pathways is modest at this horizon. By late century, however, the increase rises to 9.28\% under SSP2--4.5 and 16.88\% under SSP5--8.5, or 17.57 and 31.96 additional incidents per decade. Thus, the late-century effect under high emissions is nearly twice as large as the corresponding effect under moderate emissions. The confidence intervals exclude zero in all scenarios.

\begin{table}[htbp]
\centering
\caption{Estimated Climate-Change Impacts on Interpersonal School Shootings}
\label{tab:projected_impacts}
\begin{tabular}{llcc}
\toprule
Period & Scenario & \% Impact & U.S. School Shootings (Decadal)
 \\
\midrule
Mid-century (2051--2060) & SSP2--4.5 & 6.37 (1.57, 11.56) & 12.06 (2.96, 21.89) \\
Mid-century (2051--2060) & SSP5--8.5 & 8.49 (1.94, 15.67) & 16.07 (3.68, 29.68) \\
End-century (2091--2100) & SSP2--4.5 & 9.28 (1.99, 17.30) & 17.57 (3.77, 32.76) \\
End-century (2091--2100) & SSP5--8.5 & 16.88 (2.57, 33.42) & 31.96 (4.86, 63.29) \\
\bottomrule
\end{tabular}
\begin{tablenotes}
{\footnotesize \item \textit{Notes:} Estimates are based on the interpersonal shooting estimates from column~(1) of Table~\ref{tab:type_severity}, applied to bias-corrected CMIP6 ensemble climate data averaged across seven GCMs. 95\% confidence intervals in parentheses are constructed using the delta method. Decadal counts are computed by multiplying the percentage impact by the baseline number of interpersonal school shootings per decade. SSP2--4.5 is a moderate-emissions pathway. SSP5--8.5 is a high-emissions pathway.}
\end{tablenotes}
\end{table}

The divergence across emissions pathways is driven primarily by exposure to the top temperature bin. In the 2005--2014 historical climate baseline, roughly 14\% of days have maximum temperatures above 90$^{\circ}$F. By late century, this share rises to 27\% under SSP2--4.5 and 38\% under SSP5--8.5. In the impact calculation, these changes in bin frequencies are multiplied by the estimated coefficients from the interpersonal-shooting model. Because the above-90$^{\circ}$F bin has the largest estimated coefficient, the larger increase in above-90$^{\circ}$F days under SSP5--8.5 produces a larger estimated increase in interpersonal school shootings.

To benchmark these magnitudes against existing climate-crime estimates, we compare them with \citet{ranson2014crime}, who estimates late-century changes in violent crime under an emissions pathway broadly comparable to SSP2--4.5. \citet{ranson2014crime} estimates a 3.6\% increase in murder and a 3.1\% increase in aggravated assault. Both analyses use county-level data and binned temperature exposure, but they differ in outcome and temporal aggregation: \citet{ranson2014crime} study monthly counts of general violent crime, while our estimates are based on daily county-level variation in interpersonal school shootings. With these differences in mind, the estimated 9.28\% increase in interpersonal school shootings under SSP2--4.5 by late century is large relative to the corresponding estimates for general violent crime.

\begin{figure}
   { \centering
    \includegraphics[width=\linewidth]{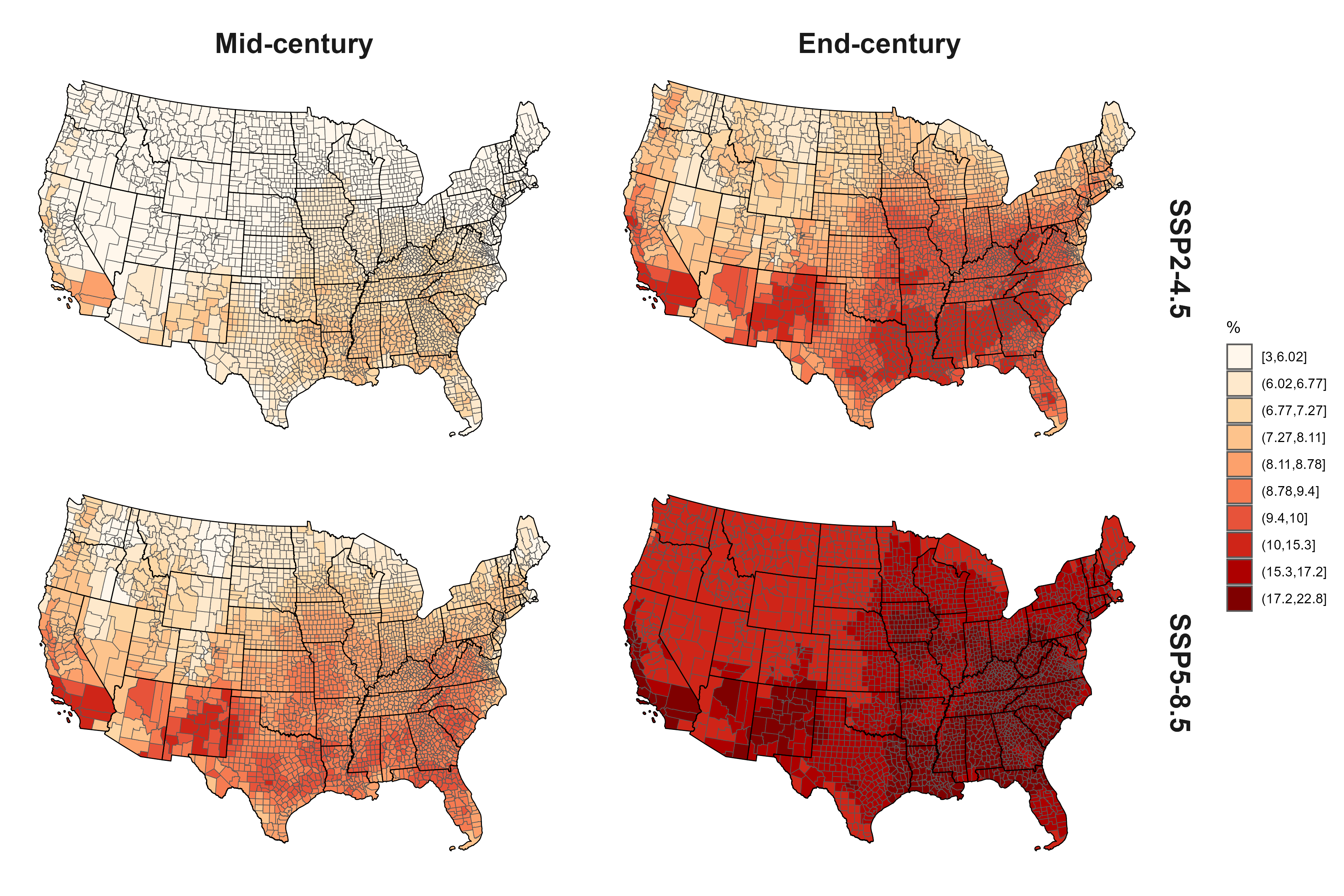}
    \caption{Estimated Climate-Change Impacts on Interpersonal School Shootings}
    \label{fig:map_prt_impact}
}
{\footnotesize \textit{Notes:} Estimates are based on historical relationships between climate variables (temperature and precipitation bins) and interpersonal school shooting incidence, applied to bias-corrected CMIP6 ensemble climate data. }
\end{figure}

Figure~\ref{fig:map_prt_impact} maps the estimated percentage impacts across counties. Under SSP2--4.5, mid-century impacts are concentrated in parts of the Southwest, Southern Plains, and Southeast. By late century, impacts become more widespread, especially across southern and central counties. Under SSP5--8.5, impacts are larger at both horizons and extend across a broader set of counties.

These estimates hold the historical temperature-response function fixed. They therefore do not incorporate long-run adaptation by schools, families, or communities. Adaptation could include cooling investments, schedule changes, increased monitoring during extreme heat, or other safety responses. If such responses reduce the sensitivity of shootings to heat, our estimates would overstate realized long-run effects. The estimates should therefore be interpreted as impacts absent substantial adaptation.

\subsubsection{Social Costs of Climate Change--Induced Interpersonal School Shootings}
We next convert estimated additional interpersonal school shootings into social cost estimates. The calculation includes three components: lost lifetime earnings among exposed students, mortality costs, and injury-related medical costs. It excludes several potentially important costs, including mental health effects on families, spillovers to teachers and communities, and institutional responses.

Using the same school-shooting dataset as in our analysis, \citet{cabral2021trauma} estimate that exposure to a school shooting is associated with a \$128{,}859 reduction (in 2025 dollars) in the present discounted value of lifetime earnings per exposed student. We assume that each incident exposes 900 students on average.\footnote{According to the \textit{Washington Post} school shooting database (\url{https://www.washingtonpost.com/education/interactive/school-shootings-database/}, accessed January~6,~2026), there have been 435 gun-violence incidents at schools since the Columbine attack in 1999, exposing more than 398{,}000 students. This implies that, on average, approximately 915 students are exposed per school shooting.} For direct harms, we use sample averages of 1.16 injuries and 0.46 fatalities per interpersonal shooting, an initial hospital care cost of \$40{,}256 per injury, and a value of a statistical life (VSL) of \$13 million per fatality.\footnote{VSL estimates follow U.S.\ Department of Transportation guidance (\url{https://www.transportation.gov/office-policy/transportation-policy/revised-departmental-guidance-on-valuation-of-a-statistical-life-in-economic-analysis}). Medical cost estimates are drawn from U.S.\ Government Accountability Office estimates (\url{https://www.gao.gov/assets/gao-21-515.pdf}).} We combine these per-incident costs with the estimated additional incidents in Table~\ref{tab:projected_impacts}: annual incidents increase by 1.2 (SSP2--4.5) and 1.6 (SSP5--8.5) during 2051--2060, and by 1.8 and 3.2, respectively, during 2091--2100.

Applying a 3\% annual discount rate, the present value of mid-century social costs is \$599 million under SSP2--4.5 and \$799 million under SSP5--8.5. Lost lifetime earnings account for over 95\% of total costs. The corresponding present values for late century are \$268 million and \$487 million. These values are lower despite larger incident counts because late-century damages are discounted over a longer horizon. Because the calculation excludes several non-earnings and community-level costs, these social cost estimates do not capture the full costs of climate-related school shooting damages.

\FloatBarrier
\section{Conclusion}
\label{sec:conclusion}

This paper estimates the effect of temperature on shootings in U.S.\ K--12 schools during regular school-day operations. Using short-run variation in local maximum temperature, we find that days with maximum temperatures above 90$^{\circ}$F increase school shooting incidence by approximately 90\% relative to days with maximum temperatures below 70$^{\circ}$F. The response is concentrated in interpersonal incidents and in less structured parts of the school day: shootings before and after class and during lunch more than triple on days with maximum temperatures above 90$^{\circ}$F, while shootings during class time show no detectable temperature response. The estimated high temperature effects are positive for both indoor and outdoor shootings and are larger for incidents involving fatalities or injuries. These patterns are consistent with heat increasing the risk of interpersonal gun violence, with timing differences suggesting that adult supervision may moderate this response.

Applying the estimated dose-response to future climate exposure, we assess impacts on interpersonal school shootings under fixed non-climate conditions and no additional adaptation. Under moderate emissions (SSP2--4.5), interpersonal shootings increase by 6\% by mid-century and 9\% by late century, or about 12 and 18 additional incidents per decade. Under high emissions (SSP5--8.5), the corresponding increases are 8\% and 17\%, or about 16 and 32 additional incidents per decade. The late-century difference across pathways reflects the larger increase in days with maximum temperatures above 90$^{\circ}$F under SSP5--8.5. These estimates are conditional impacts, not forecasts of realized outcomes under future policy, reporting, or investment responses.

The estimated additional incidents imply sizable social costs. The present discounted value of the mid-century social costs is \$599 million under SSP2--4.5 and \$799 million under SSP5--8.5. Costs to exposed students account for over 95\% of the per-incident social cost in our calculations, so the estimates are driven primarily by lost lifetime earnings rather than direct medical or mortality costs. The estimates exclude several potentially important costs, including psychological costs borne by families, teachers, and surrounding communities.

The results suggest several school-specific adaptation margins. The timing results identify arrival, dismissal, and lunch as natural margins for heat-related safety planning. Weather-related school closures in the United States currently respond mainly to snow and cold~\citep{wong2014school}. As extreme heat becomes more frequent, heat protocols may also become relevant for school safety. Investments in air conditioning, particularly in older school buildings that lack adequate climate control, could reduce heat exposure and may also reduce safety risks associated with high temperatures~\citep{park2020heat}. The paper does not estimate the returns to these interventions, but the results indicate that benefit-cost analyses of school heat adaptation should account for safety outcomes as well as learning losses. Accounting for these safety effects also changes how climate damages in schools are measured: the costs of warming may extend beyond heat stress and learning losses to include rare but high-cost safety events.

\newpage

\bibliographystyle{ecta}
\bibliography{99.bib.bib}

\newpage

 	\appendix
 	\renewcommand{\thesection}{\Alph{section}}

\counterwithin{figure}{subsection}
\counterwithin{table}{subsection}

\addcontentsline{toc}{section}{Appendix}

\section{Appendix}
\begin{figure}[!htbp]
    \centering
  
\begin{subfigure}{.9\textwidth}
  \caption{Mean Temperature ($^\circ$F)}
  \centering
  \includegraphics[width=\linewidth]{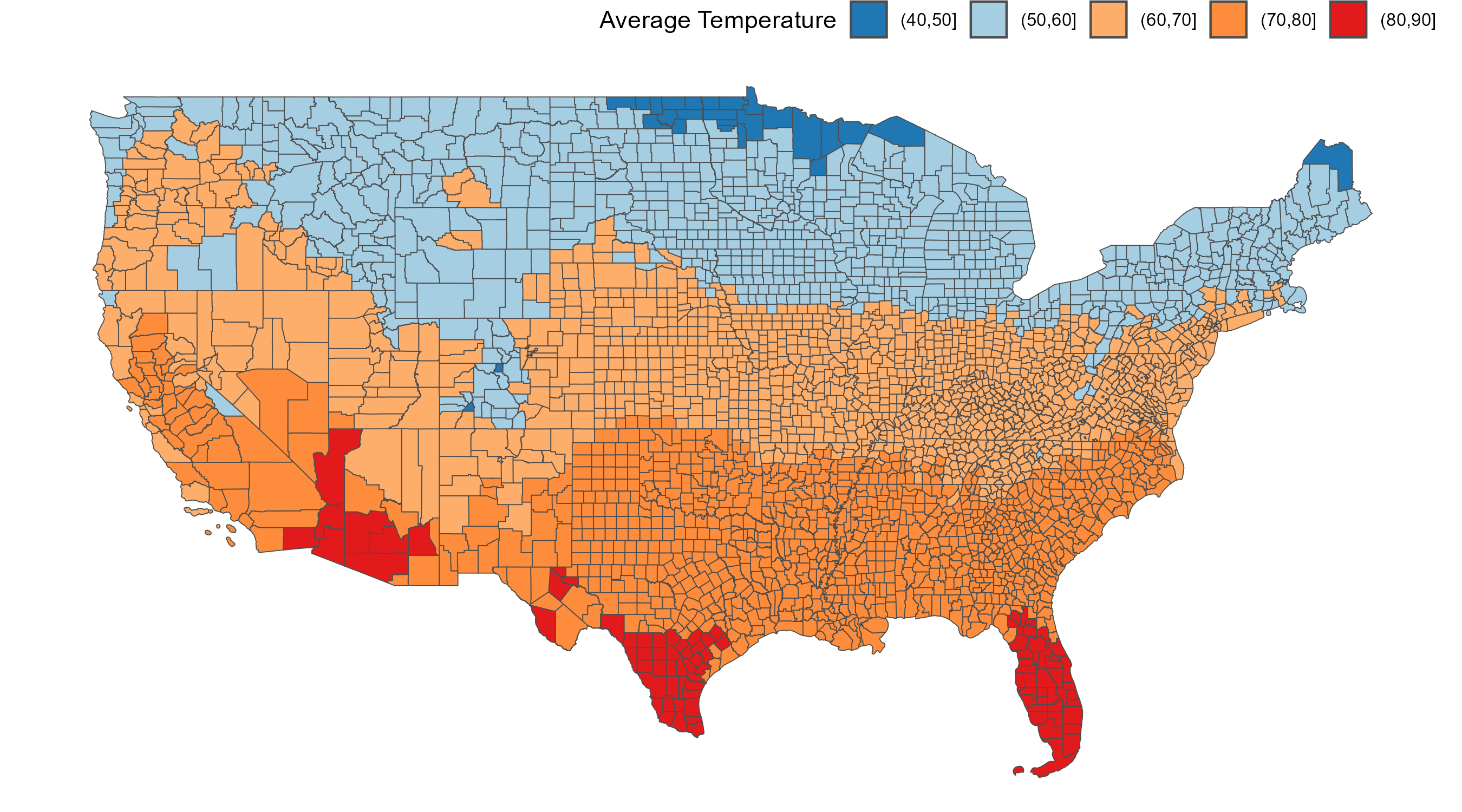} 

\end{subfigure}
\begin{subfigure}{.9\textwidth}
  \caption{Days above 90$^\circ$F}
  \centering
  \includegraphics[width=\linewidth]{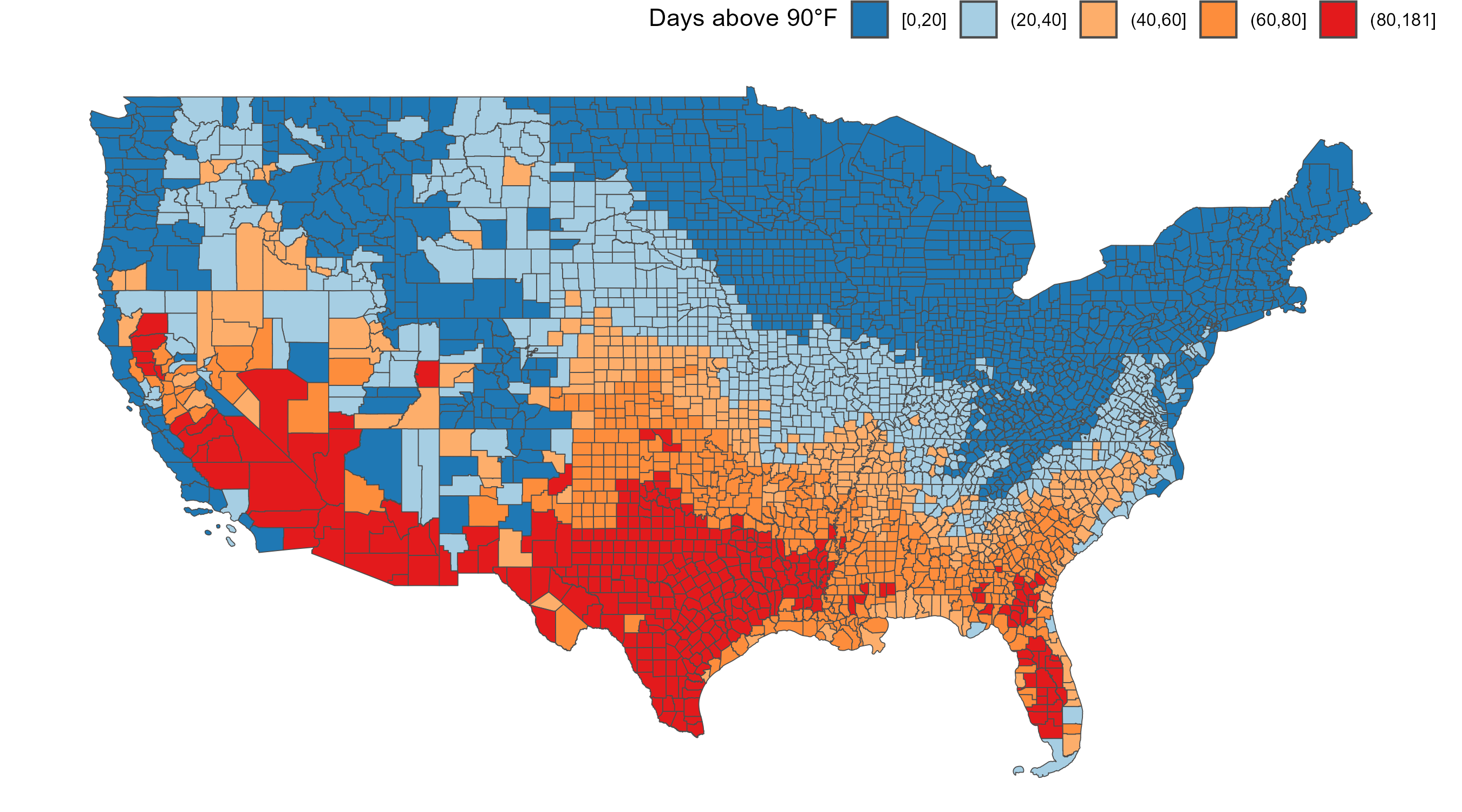} 

\end{subfigure}
    \caption{Spatial Variations of Temperature}
 \begin{minipage}{\textwidth} 
{\footnotesize \textit{Notes}: Panel (a) in Figure~\ref{fig:temp} shows the average daily maximum temperature for each county, while panel (b) displays the annual average number of days with maximum temperatures above 90$^{\circ}$F for each county in the United States.}
\end{minipage}
\label{fig:temp}
\end{figure}

\begin{figure}[!htbp]

  \centering
  \includegraphics[width=\linewidth]{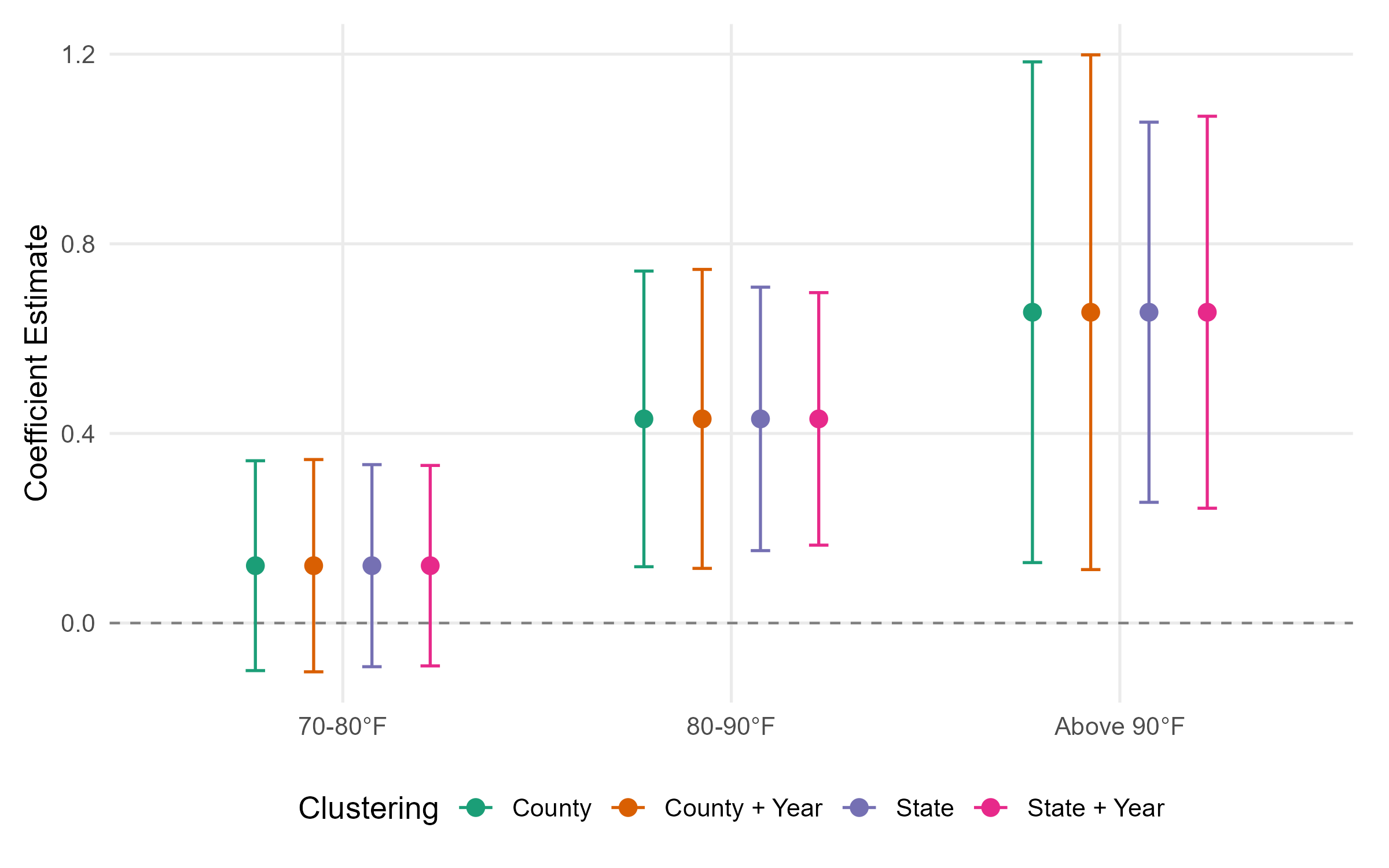}
    \caption{Robustness to Alternative Clustering Specifications}
 \begin{minipage}{\textwidth} 
{\footnotesize \textit{Notes}: The figure presents estimated coefficients following equation~\ref{eq:main}, with clustering conducted at various levels: county, county and year, state, and state and year. The x-axis represents temperature bins (in degrees Fahrenheit), while the y-axis shows the corresponding estimates of the effect.}
\end{minipage}
\label{fig:cluster}
\end{figure}

\begin{figure}
    {\centering
    \includegraphics[width=\linewidth]{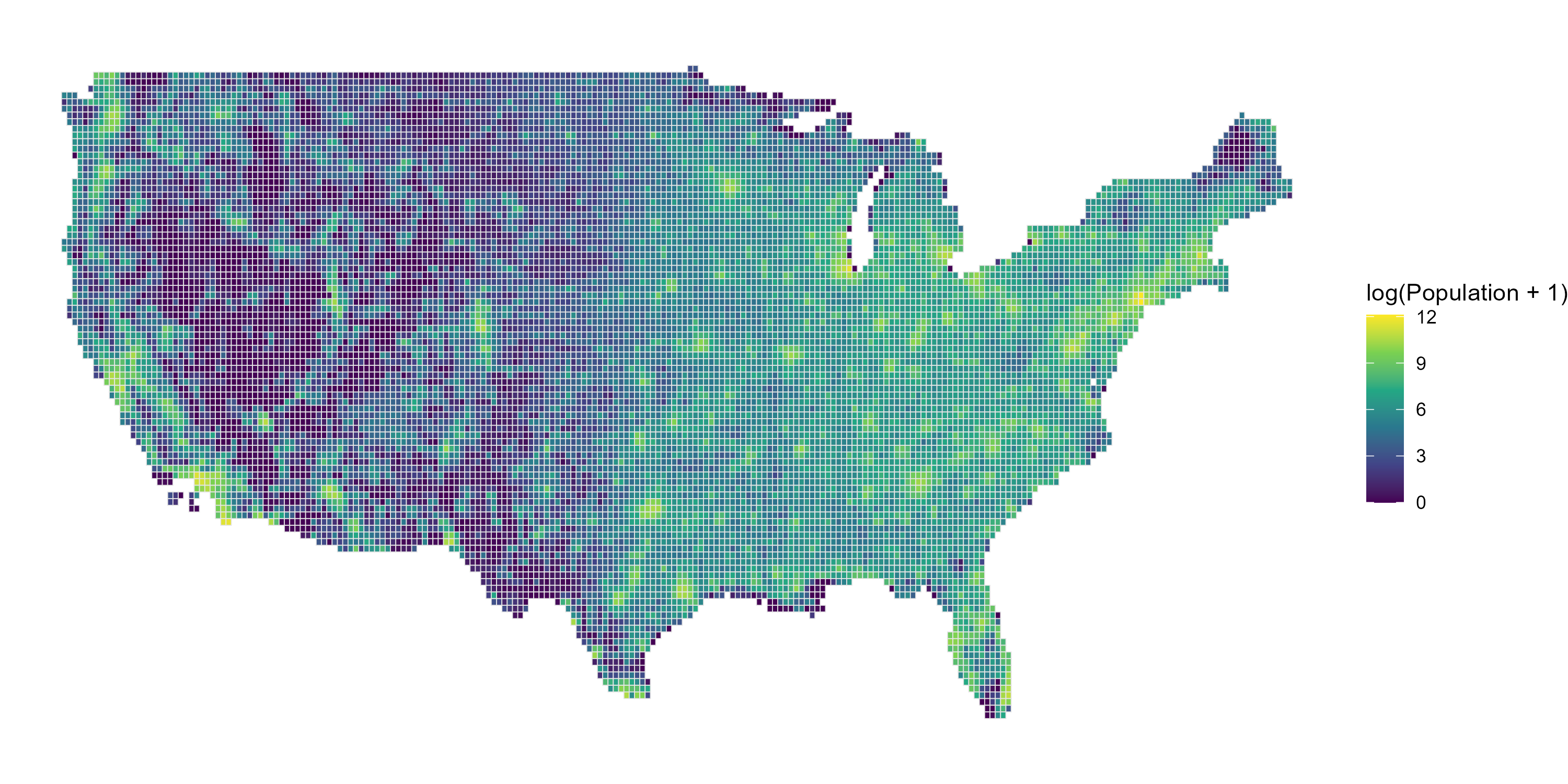}
    \caption{Population Weight for Each NEX-GDDP-CMIP6 Grid}
    \label{fig:CMIP6-weight-grid}
}
    {\footnotesize \textit{Notes}: The figure displays log(population + 1) at the native 25-km NEX-GDDP-CMIP6 resolution. Population weights are constructed by overlaying 30-m gridded population data with the NEX-GDDP-CMIP6 grid and averaging population counts within each cell. These weights are used to aggregate climate simulations to the county level.}
\end{figure}

% \begin{figure}
%    { \centering
%     \includegraphics[width=\linewidth]{Figures/Figure_files/map_prt_impact_mid.png}
%     \caption{Projected Mid-Century Impacts of Climate Change on School Shootings}
%     \label{fig:map_prt_impact_mid}
% }
% {\footnotesize \textit{Notes}: Projections are based on estimated relationships between historical climate variables (temperature and precipitation bins) and school shooting incidence, applied to bias-corrected CMIP6 ensemble climate data. }
% \end{figure}

\end{document}